\documentstyle[preprint,aps,tighten]{revtex}

\begin{document}
\draft
\title{Atomic effects in astrophysical nuclear reactions}
\author{Theodore E. Liolios $^{1,2,3}$ \footnote{theoliol@physics.auth.gr}}
\address{$^1$European Center for Theoretical Studies in Nuclear Physics and Related Areas\\
Villa Tambosi, I-38050 Villazzano (Trento), Italy\\
\footnote{Correspondence address}$^2$University of Thessaloniki, Department of Theoretical Physics\\
Thessaloniki 54006, Greece\\
$^3$ Hellenic War College, BST 903, Greece\\
}

\maketitle

\begin{abstract}
Two models are presented for the description of the electron screening
effects that appear in laboratory nuclear reactions at astrophysical
energies. The two-electron screening energy of the first model agrees very
well with the recent LUNA experimental result for the break-up reaction $%
^{3}He\left(^{3}He, 2p\right)^{4}He$, which so far defies all available
theoretical models. Moreover, multi-electron effects that enhance laboratory
reactions of the CNO cycle and other advanced nuclear burning stages, are
also studied by means of the Thomas-Fermi model, deriving analytical
formulae that establish a lower and upper limit for the associated screening
energy. The results of the second model, which show a very satisfactory
compatibility with the adiabatic approximation ones, are expected to be
particularly useful in future experiments for a more accurate determination
of the CNO astrophysical factors.
\end{abstract}


\section{Introduction}

The screening enhancement effect in laboratory nuclear reactions at
astrophysical energies has attracted a lot of attention recently, especially
after the recent accomplishments of the LUNA collaboration at Gran Sasso\cite
{lunasecond}. The very low energies attained for the break-up reaction $%
^{3}He\left( ^{3}He,2p\right) ^{4}He,$ which is extremely important to the
solar neutrino production\cite{bahcallbook}, revealed the real magnitude of
the problem, as the screening energy obtained in that experiment still
exceeds all available theoretical predictions. Other low energy experiments
of the proton-proton chain \cite{kraussdd,rolfs2000}(past, current or
planned) still need a theoretical model that could account for the observed
enhancement.

On the other hand the astrophysical factors for the reactions of the CNO
chain have been obtained by performing measurements well above the Gamow
peaks (\cite{rolfsbook} and references therein) and extrapolating to lower
energies without correcting for screening, thus committing a notable error
in certain cases as it will become apparent in this work.

Various theoretical models have been proposed so far, some of which are in
conflict with each other (e.g. accepting\cite{engstler}, or rejecting\cite
{shoppamolecular} the influence of the spectator nuclei) while others\cite
{bencze} were applied at a time when experimental measurements\cite
{krausshe3} were too sparse and inaccurate, thus their actual validity has
been obscured.

There have even been suggestions\cite{bang,langloss} that this discrepancy
between theoretical and experimental results is due to an overestimation of
the energy losses in the experiment, which cause that apparent enhancement
of the cross section. The most recent relevant experiment\cite{lunasecond}
reports no such energy-loss deficit which means that all energy losses have
been taken into account, and yet the observed screening energy is still
higher than the adiabatic limit.

Nevertheless, the prevalent belief nowadays is that a model is needed which
could give a screening energy higher than the adiabatic limit. Although the
adiabatic limit\cite{assen,bracci} is generally accepted, in a recent paper%
\cite{lioliosepj} a simple and efficient model was proposed for the study of
the screening effect on low-energy nuclear fusion reactions which exceeded
that limit. In that model, the fusing atoms were considered hydrogen-like
atoms whose electron probability density was used in Poisson's equation in
order to derive the corresponding screened Coulomb potential energy. That
way atomic excitations and deformations of the reaction participants could
also be taken into account. The derived mean-field potentials were then
treated semiclassically, by means of the WKB, in order to derive the
screening enhancement factor which was also shown to be compatible with the
experiment. In that work the screened Coulomb potentials were given without
details of their derivation. However, a detailed derivation is necessary
here before two-electron configurations are studied such as the $^{2}H\left(
^{2}H,n\right) ^{3}He$ reaction with a neutral projectile or the break-up
reaction $^{3}He\left( ^{3}He,2p\right) ^{4}He.$ This need arises from the
fact that the conventional use of a screened Coulomb potential is that of a
Yukawa one. Nevertheless, the Yukawa one is only an approximation
(truncation) of the complete screened potential arising from the solution of
Poisson's equation, as it will soon become apparent. Disregarding a priori
higher order terms can possibly induce errors, especially when the
experiment takes place at astrophysical energies of a few $keV.$

The layout of the paper is as follows: In Sec. II the screened Coulomb
potentials for hydrogen-like atoms are derived in a detailed fashion, which
turns out to be very useful in Sec. III, where those potentials are modified
in order to account for two-electron effects in nuclear reactions at
astrophysical energies. Notably, in Sec. III, the screening energy for the $%
^{3}He\left( ^{3}He,2p\right) ^{4}He$ reaction which so far remains
inexplicably\cite{lunafirst} high is reproduced to a very good
approximation. Sec IV deals with multi-electron effects by means of the
Thomas-Fermi model, which enables us to derive analytical formulas for the
screening enhancement factors for reactions encountered in advanced nuclear
burning stages of stellar evolution. A final concluding section summarizes
the novelties and the results of the present work.

\section{One-electron screening effects}

Let us consider a hydrogen-like atom with atomic number $Z_{1}$. When the
wave function of the electron is given by $\Psi _{nl}\left( r,\theta \right) 
$ then the charge density around the point-like nucleus is

\begin{equation}
\rho _{nl}\left( r,\theta \right) =-e\left| \Psi _{nl}\left( r,\theta
\right) \right| ^{2}  \label{density}
\end{equation}
Assuming spherically symmetric wavefunctions for simplicity we can solve the
equation of Poisson for the above charge density in order to derive a
screened Coulomb potential $\Phi \left( r\right) $ around the nucleus. Note
that this potential will take into account the repulsive effects of the
point-like nucleus, by imposing the appropriate boundary conditions:

\begin{equation}
\Phi \left( \infty \right) =0\qquad ,\qquad \Phi \left( 0\right) =\frac{%
Z_{1}e}{r}
\end{equation}
The second boundary condition indicates that if a positive projectile $%
\left( Z_{2}e\right) $ is in contact with the nucleus $\left( Z_{1}e\right) $
at the center of the electron cloud then there is no negative charge between
them to reduce the Coulomb barrier.

Let us define the screening form factor of the screened Coulomb as a
function $f\left( r\right) $ so that:

\begin{equation}
\Phi \left( r\right) =\frac{Z_{1}e}{r}f\left( r\right)  \label{ff}
\end{equation}
If we insert Eq.$\left( \ref{ff}\right) $ and Eq. $\left( \ref{density}%
\right) $ into the equation of Poisson we obtain

\begin{equation}
\frac{d^{2}f_{nl}\left( r\right) }{dr^{2}}=\frac{4\pi }{Z_{1}}r\left| \Psi
_{nl}\left( r\right) \right| ^{2}  \label{f2k}
\end{equation}
which is to be solved with the boundary conditions

\begin{equation}
f_{nl}\left( \infty \right) =finite.\qquad ,\qquad f_{nl}\left( 0\right) =1
\end{equation}
Equation $\left( \ref{f2k}\right) $ with the corresponding boundary
conditions constitutes a generator of screened Coulomb potentials which
correspond to a particular excitation and ionization of the atomic target.
According to the quantum state of the hydrogenlike atom we can use Eq. $%
\left( \ref{f2k}\right) $ in order to obtain the corresponding screened
Coulomb potentials.

Actually in the above treatment there is an implicit assumption of
independence between the nuclear and electronic degrees of freedom. This
assumption can be expressed in a quantitative form by the formula:

\begin{equation}
V_{sc}\left( r\right) =V_{c}\left( r\right) +\Phi _{e}\left( r\right)
\end{equation}
The above formula states that the screened Coulomb potential is actually the
sum of the bare nucleus Coulomb potential $V_{c}\left( r\right) $ plus the
electrostatic potential energy of the electron cloud. Therefore one can
calculate $\Phi _{e}\left( r\right) $ and then add it to the bare nucleus
potential which simplifies the calculations especially when multi-electron
atoms are considered. We now proceed to give an alternative approach
according to the above assumption, which will be particularly useful in our
study of two-electron effects.

Let us assume a hydrogenlike atom $Z_{1}$ in its ground state, whose
electron charge distribution of Eq. $\left( \ref{density}\right) $ can be
written:

\begin{equation}
\rho \left( r\right) =\rho \left( 0\right) \exp \left( -r/r_{0}\right)
\end{equation}
where $r_{0}=a_{0}/\left( 2Z_{1}\right) $.

The charge density at the center of the electron cloud is:

\begin{equation}
\rho \left( 0\right) =-\frac{e}{\pi }\left( \frac{Z_{1}}{a_{0}}\right) ^{3}
\end{equation}
while the electrostatic potential of the distribution is given by the
solution of Poisson's equation: 
\begin{equation}
\frac{1}{r^{2}}\frac{d}{dr}\left( r^{2}\frac{d\Phi _{e}\left( r\right) }{dr}%
\right) =-4\pi \rho \left( 0\right) \exp \left( -r/r_{0}\right)
\end{equation}
Upon integration we obtain :

\begin{equation}
\Phi _{e}\left( r\right) =C_{1}+\frac{C_{2}}{r}-4\pi \rho \left( 0\right)
r_{0}^{2}\exp \left( -\frac{r}{r_{0}}\right) \left( 1+2\frac{r_{0}}{r}\right)
\end{equation}
The electrostatic potential $\Phi _{e}\left( r\right) $ must go to zero at
infinity which gives $C_{1}=0.$ At very large distances $r>>r_{0}$, due to
the spherical symmetry of the distribution , any projectile impinging on
that cloud will actually ''see'' a Coulomb potential of the form:

\begin{equation}
\Phi _{e}\left( r>>r_{0}\right) =\frac{-e}{r}
\end{equation}
so that $C_{2}=-e.$ Inserting the values of the parameters $r_{0}$ and $\rho
\left( 0\right) \,$into the above equation we obtain the formula used
without details of its derivation in Ref. \cite{lioliosepj}:

\begin{equation}
\Phi _{e}\left( r\right) =-\frac{e}{r}+\frac{e}{r}\left( 1+\frac{r}{2r_{0}}%
\right) \exp \left( -r/r_{0}\right)  \label{fe}
\end{equation}
If a positive projectile $Z_{2}e$ impinges onto the above hydrogen-like atom
the total interaction potential energy $V\left( r\right) $ between the two
nuclei will be due to the above electrostatic potential, that is $Z_{2}e\Phi
_{e},$ plus the repulsive potential of the nucleus $Z_{1}e:$

\begin{equation}
V_{sc}\left( r\right) =\frac{Z_{1}Z_{2}e^{2}}{r}-\frac{Z_{2}e^{2}}{r}+\frac{%
Z_{2}e^{2}}{r}\left( 1+\frac{r}{2r_{0}^{*}}\right) \exp \left(
-r/r_{0}^{*}\right)   \label{v12}
\end{equation}
where

\begin{equation}
r_{0}^{*}=\frac{a_{0}}{2\left( Z_{1}+Z_{2}\right) }
\end{equation}
The reason for replacing $r_{0}$ with $r_{0}^{*}$ is that, at astrophysical
energies, the electrons move at higher velocities than the nuclei
themselves. For example in laboratory $d-D$ reactions the relative nuclear
velocity equals the typical electron velocity $v_{e}=\alpha c$ for $E=25keV.$
Although the above assumption is particularly valid at such low energy
collisions between hydrogen nuclei, when reactions between heavier nuclei
are considered (see next section), an inevitable small error is involved at
intermediate energies (e.g. in the vicinity of the Gamow peak). As it will
soon become clear the WKB treatment of the penetration factor disregards all
effects beyond the classical turning point. Therefore, inside the tunneling
region, the wavefunction of the electron actually corresponds to a combined
nuclear molecule $\left( Z_{1}+Z_{2}\right) $ instead of the initial $Z_{1}$
atom. Of course this is an approximation and, in our model, the intermediate
stages of the wavefunction deformations are assumed to play a minor role.

When the electron of the target atom/ion is in an excited state, we can
obtain in the same way the corresponding potential energy used in Ref.\cite
{lioliosepj}.

\section{Two-electron screening effects}

Let us now assume that a hydrogen-like atom $Z_{2}e$ impinges on a resting
target nucleus $Z_{1}e.$ Let us further assume that the target atom has two
electrons orbiting the nucleus. In a Hartree-Fock approximation the total
potential energy of the interaction will be: 
\begin{equation}
V_{sc}\left( r\right) =V_{c}\left( r\right) +V_{n_{2}e_{1}}\left( r\right)
+V_{n_{2}e_{2}}\left( r\right) +V_{n_{1}e}\left( r\right)
+V_{e_{1}e_{2}}+V_{e_{1}e}+V_{e_{2}e}
\end{equation}
that is the sum of: a) the Coulomb potential energy $V_{c}\left( r\right) $
between the two bare nuclei plus b) the interaction between the projectile $%
\left( n_{2}\right) $ and the electrons $\left( e_{1,2}\right) $ of the
target nucleus, plus c) the interaction between the target nucleus $\left(
n_{1}\right) $ and the electron of the projectile $\left( e\right) ,$plus d)
the interaction between the electrons of the target $V_{e_{1}e_{2}}$ and of
course , the interaction $\left( e_{1}e,e_{2}e\right) \,$between the
electron of the projectile and those of the target. In the above equation
only the terms associated with the nuclei will be considered functions of
the relative internuclear distance, while the electron-electron interactions
will be treated as perturbations which will actually raise the Coulomb
barrier between the two reacting nuclei. We consider the following channels:

a) The nucleus-nucleus channel

That interaction has been thoroughly studied in most text books\cite
{rolfsbook} and needs no further elaboration.

b) The atom-atom channel for hydrogen like atoms.

In most experiments, the projectile has been considered fully stripped of
its electrons which is the case at relatively high energies. However, when
the projectile is in a neutral state, or at least not fully ionized, its
electron cloud has to be taken into account\cite{engstler}.

For such an interaction the total potential energy can be written:

\begin{equation}
V_{sc}\left( r\right) =V_{c}\left( r\right) +V_{n_{2}e_{1}}\left( r\right)
+V_{n_{1}e}\left( r\right) +V_{e_{1}e}  \label{vaa}
\end{equation}
When the two electron clouds interact, their mutual ground state
wavefunction must be antisymmetric, since the electrons are identical
fermions. The spatial wavefunction is necessarily symmetric therefore
antisymmetry is arranged by considering an antisymmetric spin singlet state.
Since we still work with hydrogen-like atoms, the electron-electron spatial
wavefunction is

\begin{equation}
\Psi _{e_{1}e}\left( \overrightarrow{r}_{1},\overrightarrow{r}_{2}\right)
=\Psi _{00}\left( \overrightarrow{r}_{1}\right) \Psi _{00}\left( 
\overrightarrow{r}_{2}\right)
\end{equation}
where $\Psi _{00}\left( \overrightarrow{r}\right) $ is actually the usual
ground state wavefunction of hydrogen-like atoms$.$

The electrostatic potential energy of the two electrons is:

\begin{equation}
V_{e_{1}e}=\int_{1}\int_{2}\frac{e^{2}}{\left| \overrightarrow{r}_{1}-%
\overrightarrow{r}_{2}\right| }\left| \Psi _{00}\left( \overrightarrow{r}%
_{1}\right) \right| ^{2}\left| \Psi _{00}\left( \overrightarrow{r}%
_{2}\right) \right| ^{2}d^{3}r_{1}d^{3}r_{2}
\end{equation}
where $\overrightarrow{r}_{1,2}$ are the positions of the two electrons. As
it will soon become apparent the effective interaction between the two
participants of the reaction begins upon reaching the classical turning
point. At astrophysical energies the classical turning point is hundreds of
times smaller than the atomic radius, therefore we can safely assume that
throughout the Coulomb barrier the two colliding nuclei constitute a
combined nuclear molecule. Under that assumption the above integral can be
calculated \cite{gazbook} so that for $Z_{1}=Z_{2}=Z$ we obtain:

\begin{equation}
V_{ee}=\frac{5}{8}\frac{Ze^{2}}{a_{0}}
\end{equation}
That positive energy will be transferred to the relative nuclear motion,
increasing the height of the Coulomb barrier.

The equal charge assumption combined with the fact that during tunneling the
reacting nuclei practically coincide with respect to the electron cloud
dimensions yields: $V_{n_{2}e_{1}}\left( r\right) \simeq V_{n_{1}e}\left(
r\right) $

On the other hand each of the two electrons is actually subject to the
repulsive effect of a screened nucleus due to the presence of the other
electron. For the combined nuclear molecule, we have $Z_{t}=Z_{1}+Z_{2}$ ,
while the usual variational procedure yields an effective atomic number for
each electron $Z^{**}=Z_{t}-5/16.$

Therefore for the low energy reaction of two hydrogenlike atoms in their
ground state, with equal atomic numbers $Z$, the interaction potential
energy is :

\begin{equation}
V_{sc}\left( r\right) =\frac{Z^{2}e^{2}}{r}-2\left[ \frac{Ze^{2}}{r}+\frac{%
Ze^{2}}{r}\left( 1+\frac{r}{2r_{0}^{**}}\right) \exp \left(
-r/r_{0}^{**}\right) \right] +\frac{5}{8}\frac{Z^{**}e^{2}}{a_{0}}
\label{vsc2}
\end{equation}
where

\begin{equation}
r_{0}^{**}=\frac{a_{0}}{2Z^{**}}
\end{equation}
In the same way we can calculate the potential energy when one (or both)
atoms are in an excited state.

c) The nucleus-atom channel.

For the hydrogen-like atom target an extensive study has already been given%
\cite{lioliosepj}. However, for a two electron target atom further
elaboration is needed. In that case the potential energy is :

\begin{equation}
V\left( r\right) =V_{c}\left( r\right) +V_{n_{2}e_{1}}\left( r\right)
+V_{n_{2}e_{2}}\left( r\right) +V_{e_{1}e_{2}}  \label{vna}
\end{equation}
If we compare Eq. $\left( \ref{vaa}\right) $ with Eq.$\left( \ref{vna}%
\right) $ we see that the two potentials must be approximately the same.
Therefore the potential energy $\left( \ref{vsc2}\right) $ can account for
the nucleus-atom channel with a two-electron target atom, as well.

The penetration factor $P\left( E\right) $ multiplied by the astrophysical
factor $S\left( E\right) $ in the $s$-wave cross section formula

\begin{equation}
\sigma \left( E\right) =\frac{S\left( E\right) }{E}P\left( E\right)
\end{equation}
is given by the WKB method:

\begin{equation}
P\left( E\right) =\exp \left[ -\frac{2\sqrt{2\mu }}{\hbar }%
\int_{R}^{r_{c}\left( E\right) }\sqrt{V_{sc}\left( r\right) -E}dr\right]
\label{pe}
\end{equation}
where the classical turning point is given by

\begin{equation}
V_{sc}\left( r_{c}\right) =E  \label{rc}
\end{equation}
At astrophysical energies the potential energy is found to be shifted by a
constant screening energy $U_{e}$ which is added to the relative energy of
the collision.

For the nucleus-nucleus channel the calculation is trivial leading to $%
P\left( E\right) =\exp \left( -2\pi n\right) $ where $n$ is the Sommerfeld
parameter and $U_{e}=0$.

For the nucleus-hydrogenlike atom channel the screening energy has already
been calculated \cite{lioliosepj} 
\begin{equation}
U_{e}=-\left( Z_{1}+Z_{2}\right) Z_{2}\frac{e^{2}}{a_{0}}  \label{usl}
\end{equation}
For two hydrogenlike atoms with equal charges $\left( Ze\right) $ we have
performed numerical integrations of Eq. $\left( \ref{pe}\right) $ and
numerical solutions of Eq. $\left( \ref{rc}\right) $. At astrophysical
energies where screening becomes important the screened coulomb potential
can be safely replaced by the quantity

\begin{equation}
V_{sc}\left( r\right) =\frac{Z^{2}e^{2}}{r}+U_{e}
\end{equation}
where:

\begin{equation}
U_{e}=-2Z\left( 2Z-5/16\right) \frac{e^{2}}{a_{0}}+\frac{5}{8}\left(
2Z-5/16\right) \frac{e^{2}}{a_{0}}  \label{u2}
\end{equation}
which is also the screening energy for the collision of a bare nucleus $%
\left( Ze\right) $ with a two-electron target atom.

For the astrophysical reaction $^{3}He\left( ^{3}He,2p\right) ^{4}He$ we
have $Z_{1}=Z_{2}=Z=2$ and the corresponding screening energy obtained
through the above model is $U_{e}=-338$ $eV$. Our result is very close to
the experimental result of the LUNA collaboration\cite{lunasecond} $%
U_{e}^{ex}=-294\pm 47\,eV.\,$The small difference could be plausibly
attributed to energy losses and experimental errors had it not been for
another, as yet unidentified, uncertainty source for the screening energies
and the associated astrophysical factors, which causes a small increase in
the uncertainty for the solar neutrino fluxes. In fact the LUNA experimental
results were first\cite{lunafirst} fitted using three different approaches:
a) fixing the screening energy at the value given by the adiabatic limit, b)
allowing all three parameters $\left( S\left( 0\right) ,S^{^{\prime }}\left(
0\right) ,S^{^{\prime \prime }}\left( 0\right) \right) $ of the
astrophysical factor and the screening energy $U_{e}$ to vary
simultaneously, c) using higher energy data to fix the parameters of the
astrophysical factor while varying the screening energy. The two model
independent methods (b,c) gave considerably different screening energies in
that work, that is $U_{e}^{\left( b\right) }=-323\pm 51$ $eV$ and $%
U_{e}^{\left( c\right) }=-432\pm 29$ respectively. In their final article%
\cite{lunasecond} only the $\left( b\right) $ method was used yielding the
above mentioned value of $U_{e}^{ex}=294\pm 47\,eV$. It is obvious that if
the (c) method had been used the screening energy would have been higher,
and the adiabatic limit would have been considerably exceeded again. The two
methods (b,c) gave respectively \cite{lunafirst} the following zero energy
astrophysical factors: $S^{\left( b\right) }\left( 0\right) =5.30\pm 0.08$
and $S^{\left( c\right) }\left( 0\right) =5.1\pm 0.1,$ that is a $3.9\%$
difference. Regarding the solar neutrino problem that percentage admittedly
leads to a negligible neutrino flux uncertainty for the $pp$ and $hep$
neutrinos ($0.1\%).$ However the uncertainty for the $^{8}B$ and $^{7}Be$
ones can be as high as $1.5\%$ , which should not be disregarded. Although
the author agrees that method (b) seems more plausible than the (c) one, a
better justification is needed for the choice of the fitting method when
such low energy experiments are considered as it obviously constitutes a
source of uncertainty.

On the other hand for the reaction $^{2}H\left( d,p\right) ^{3}H$ , where
the projectile is considered neutral and the target is in an atomic form we
have $\;Z=1,$ therefore (Eq. $\left( \ref{u2}\right) $): $U_{e}=\allowbreak
-63$ $eV.$ If we compare that screening energy with the corresponding one
for a bare projectile deuteron \cite{lioliosepj}$\left( U_{e}=-54\right) $
we arrive at an anticipated result. The screening energy is indeed higher in
the neutral projectile case but not twice as much as that of the bare
nucleus case. That is due to the repulsive effect between the two electrons
which is roughly $V_{ee}=+29\,\,eV$ and the screening of the combined
nuclear molecule by the second electron which of course yields an effective
charge which is lower than the sum of the two separate nuclei and
consequently a sparser charge distribution.

\section{Multi-electron screening effects}

In the framework of the Thomas-Fermi (TF) model the screened Coulomb
potential for a neutral multi-electron atom $Z_{1}$ is\cite{flugge} :

\begin{equation}
V_{sc}^{TF}\left( r\right) =\frac{Z_{1}e}{r}\phi \left( r\right)  \label{tf}
\end{equation}
The dimensionless function $\phi \left( r\right) $ can be obtained by the
solution of the universal differential equation

\begin{equation}
\frac{d^{2}\phi \left( x\right) }{dx^{2}}=\frac{\phi ^{3/2}\left( x\right) }{%
\sqrt{x}}
\end{equation}
where $x=r/a$ and the screening radius is $a=0.8853Z_{1}^{-1/3}a_{0}.$

In our approach, instead of solving numerically the differential equation,
we will make use of an analytic approximation of the function $\phi \left(
r\right) $ given by Tietz\cite{tietz}. Namely:

\begin{equation}
\phi \left( x\right) =\frac{1}{\left( 1+bx\right) ^{2}}  \label{tietz}
\end{equation}
with $b=0.536.$

At first we will consider the interaction of very light nuclei with a heavy
multi-electron atom so that we can disregard the perturbation induced by the
impinging particle to the average electron density of the target atom, which
is the sudden limit (SL) approximation.

If we expand the TF screened potential using Tietz's approximation we obtain
the screened potential energy of the interaction:

\begin{equation}
V_{sc}^{TF}=\frac{Z_{1}Z_{2}e^{2}}{r}-2\frac{bZ_{1}Z_{2}e^{2}}{a}+3\frac{%
Z_{1}Z_{2}b^{2}e}{a^{2}}r+O\left( r^{2}\right)   \label{tscr}
\end{equation}
The third term of the above potential, as well as terms O$\left(
r^{2}\right) ,\,$are negligible with respect to the constant screening
energy shift given by the second term. For example for relative energies of $%
20\,keV$ the tunneling region for a $p+^{14}N$ reaction begins at a
classical turning point of $r_{c}=500fm.$ At such a distance the second term
is $440eV\,$while the third is only $7eV.$ Therefore, inside the barrier,
where $r<r_{c}$, terms proportional to the scaled distance $r/a$ practically
vanish.

Inserting the above screened Coulomb potential $\left( \ref{tscr}\right) \,$%
with Tietz's approximation into the WKB integral of Eq. $\left( \ref{pe}%
\right) \,$and working in the same way as with the potentials of the
previous section we obtain the screening energy for the collision of a light
bare nucleus $\left( Z_{2}e\right) \,$with a neutral multi-electron atom $%
\left( Z_{1}e\right) $:

\begin{equation}
U_{TF}^{SL}=-1.21Z_{1}^{4/3}Z_{2}\frac{e^{2}}{a_{0}}  \label{utfe}
\end{equation}
Let us now suppose that the impinging nucleus (not necessarily a light one)
has been neutralized so that it can also be considered a TF atom. We can
obtain the maximum screening energy transferred to the relative nuclear
motion by using the formula for the total energy of a TF atom. In fact there
are three contributions to the total energy of the atom: The kinetic energy
of the electrons, the potential energy of their interaction with the nucleus
and the potential energy of their mutual interaction. For a neutral TF atom
with charge $Z_{1}e$ the total energy is given by\cite{flugge}:

\begin{equation}
E_{TF}^{tot}=-\frac{Z_{1}^{2}e^{2}}{a}\left( \frac{2}{5}\mu -\frac{J}{10}%
\right)
\end{equation}
where $\mu =-\phi ^{^{\prime }}\left( 0\right) $ and

\begin{equation}
J=\int_{0}^{\infty }\left( \frac{d\phi }{dx}\right) ^{2}dx
\end{equation}
Numerically,

\begin{equation}
E_{TF}^{tot}=-20.93\,Z_{1}^{7/3}\,eV
\end{equation}
It is therefore plausible to assume that for an adiabatic limit (AL)
interaction of two neutral TF atoms the screening energy $U_{TF}^{AL}$ will
be the difference between the total energy of the combined molecule and that
of the two separate atoms:

\begin{equation}
U_{TF}^{AL}=-20.93\left[ \left( Z_{1}+Z_{2}\right)
^{7/3}-Z_{1}^{7/3}-Z_{2}^{7/3}\right] \text{ }eV  \label{utfs}
\end{equation}
We can compare the results of Eqs. $\left( \ref{utfe}\right) \,$and $\left( 
\ref{utfs}\right) $ with the ones obtained in Ref.\cite{bracci}$.$ For a
reaction $p+_{Z}^{A}X$ between a bare proton $p$ and neutral atom $_{Z}^{A}X$
we obtain:

\bigskip

{\bf Table I. The screening energies for various proton-induced
astrophysical reactions as obtained through the sudden $\left(
U_{e}^{SL}\right) $ and the adiabatic $\left( U_{e}^{AL}\right)$ limit of
Ref. \cite{bracci} versus the screening energies obtained through the
present TF sudden $\left( U_{TF}^{SL}\right)$ and adiabatic $%
\left(U_{TF}^{AL}\right)$ limit.}

\begin{center}
$
\begin{array}{llllll}
& p+_{3}^{7}L & p+_{5}^{11}B & p+_{6}^{12}C & p+_{7}^{14}N & p+_{8}^{18}O \\ 
U_{e}^{AL}\,\left( eV\right)  & 186 & 347 & 441 & 544 & 653 \\ 
U_{e}^{SL}\,\left( eV\right)  & 134 & 281 & 366 & 462 & 570 \\ 
U_{TF}^{SL}\,\left( eV\right)  & 142 & 281 & 359 & 441 & 527 \\ 
U_{TF}^{AL}\,\left( eV\right)  & 259 & 474 & 592 & 717 & 847
\end{array}
$
\end{center}

\bigskip

We observe that Eq. $\left( \ref{utfe}\right) $ practically reproduces the
sudden limit $U_{e}^{SL}$ of Ref.\cite{bracci}$,$ while Eq. $\left( \ref
{utfs}\right) $ gives a higher adiabatic limit than the one $\left(
U_{e}^{AL}\right) $ given in Ref. \cite{bracci}$.$ Despite the fact that for
the reaction $^{3}He\left( ^{3}He,2p\right) ^{4}He\,$the electrons involved
are too few to justify use of the above formulas it will give a sense of the
validity of the present models if we apply our formulas on that reaction as
well. In fact Eq. $\left( \ref{utfe}\right) $ gives $U_{TF}^{SL}=166$ $eV$
while Eq. $\left( \ref{utfs}\right) $ gives $U_{TF}^{AL}=426$ $eV,$ which
are admittedly reasonable bounds.

Regarding the acceleration effects produced by the above screening energies
on the non-resonant wings of astrophysical nuclear reactions , we can apply
the usual enhancement factor\cite{shoppaatomic}

\begin{equation}
f_{TF}^{\left( SL,AL\right) }\left( E\right) =\exp \left( \pi n\frac{%
U_{TF}^{\left( SL,AL\right) }}{E}\right)
\end{equation}
where $n$ is the Sommerfeld parameter and $E$ is the center-of-mass energy.
We obtain for the sudden and the adiabatic limit respectively :

\begin{equation}
f_{TF}^{SL}\left( E\right) \simeq \exp \left( \frac{%
Z_{1}^{7/3}Z_{2}^{2}A^{1/2}}{2E_{\left( keV\right) }^{3/2}}\right)
\label{fsl}
\end{equation}
and

\begin{equation}
f_{TF}^{AL}\left( E\right) \simeq \exp \left[ \frac{Z_{1}Z_{2}\left[ \left(
Z_{1}+Z_{2}\right) ^{7/3}-Z_{1}^{7/3}\right] A^{1/2}}{3E_{\left( keV\right)
}^{3/2}}\right]  \label{fal}
\end{equation}
where $A=A_{1}A_{2}\left( A_{1}+A_{2}\right) ^{-1}$ is the reduced mass
number.

The above formula will be particularly useful in laboratory experiments
where such nuclear reactions are involved as those of the CNO cycle and the
ones encountered in the final stages of stellar evolution (e.g. supernova
nucleosynthesis). In such reactions , it is often impossible to measure a
nonresonant cross section in the laboratory at a sufficiently low energy due
to technical difficulties\cite{rolfsbook} (large background of counts, beam
instability etc.). Therefore the associated astrophysical factor cannot be
accurately extrapolated to zero energies as required for the calculation of
the effective astrophysical factor $S_{eff}$ that appears in the
thermonuclear reaction rate formulas (Ref. \cite{lioliosprc} and references
therein). However, the recent accomplishments of the LUNA collaboration with
the $^{3}He\left( ^{3}He,2p\right) ^{4}He$ reaction inspire hope that
similar low-energy experiments will soon be conducted for heavier nuclei
(multi-electron atoms) as well, in which case the above formula will help
correct the low energy cross section measurements.

Let us consider for example the first member of the $CNO$ cycle, namely the
radiative direct capture reaction $^{12}C\left( p,\gamma \right) ^{13}N$.
The low energy cross section is dominated by an s-wave resonance $\left( 
\frac{1}{2}^{+}\right) $ at $E_{cm}=424$ $keV,$ while its Gamow peak at
central solar conditions is $24.5$ $keV.$ The usually employed $S_{eff}$ for
that reaction is the one obtained by an experiment\cite{azuma} which
measured cross sections at energies as low as $E_{cm}=138\,keV$ disregarding
all screening effects. According to the present models the screening
enhancement of the cross section at such an energy would be between $2\%$
and $3.4\%$ which is not a negligible correction. It is now obvious that in
any future attempt to improve the accuracy of the extrapolation by lowering
the energy of the experiment the proposed model of this work will be very
useful.

The corrections are even more important for other reaction of the CNO cycle
such as the slowest one, which controls the energy generation of that cycle,
that is $^{14}N\left( p,\gamma \right) ^{15}O$. Various investigations\cite
{schroeder} (and references therein) have obtained data for that reaction to
center-of-mass energies as low as $93$ $keV$ without correcting for
screening. The error committed at such low energies for that particular
reaction can be as high as $9\%$ (AL). Finally, the corrections can be
dramatic if we consider very low resonances such as the $66$ $keV$ one of
the $^{17}O\left( p,a\right) ^{14}N$ reaction, namely of the order of $12\%$ 
$\left( AL\right) $. Such considerable enhancement could presumably lead
even to shifts of the resonance energies themselves, compared to the bare
nuclei measurements.

In figures 1 and 2, there is plotted the screening enhancement factor for
the most important astrophysical nuclear reactions of the $CNO$ bi-cycle
with respect to the center-of-mass energy. Close to the Gamow peaks of those
reactions (Fig.1) the predicted enhancement of the nonresonant cross section
is already $5\%$ (at least) while at very low energies(Fig.2) $2\leq
E_{cm}\leq 5keV$, the nonresonant cross section of the screened reaction is
expected to be several times larger than that of the bare nucleus one.

Before concluding this section we should emphasize that our results agree
well with the ones obtained by the adiabatic assumption\cite{assen}
according to which the screening energy of the reaction between atoms $A$
and $B$ is given by the formula:

\begin{equation}
U_{e}^{BE}=E_{el}\left( A+B\right) -E_{el}\left( A\right) -E_{el}\left(
B\right) 
\end{equation}
where $E_{el}\left( A+B\right) $ is the total binding energy of the
electrons in the combined atom $A+B$ , and $E_{el}\left( A\right)
,\,E_{el}\left( B\right) $ are the total electronic binding energies of the
asymptotically separated atomic reaction partners $A,B.$ For example for the 
$^{12}C\left( p,\gamma \right) ^{13}N$ and the $^{14}N\left( p,\gamma
\right) ^{15}O$ reactions respectively, the latter model gives screening
energies $U_{e}^{BE}=444$ $eV$ and $546$ $eV$, which are bracketed by the
energies derived through the present models (see table)

\section{Conclusions}

Two very efficient models are presented here which reproduce the screening
enhancement effects that appear in laboratory nuclear reactions at
astrophysical energies. The first model, which describes two-electron
effects, relies on the Hartree-Fock approximation and agrees very well with
the \ recent LUNA experimental screening energy for the reaction $%
^{3}He\left( ^{3}He,2p\right) ^{4}He$, which so far remains unexplained.

The second model is based on the Thomas-Fermi theory and yields the
screening energies for reactions encountered in advanced nuclear burning
stages of stellar evolution, where multi-electron effects dominate. To the
author's knowledge, for multi-electron laboratory effects, there have never
been any closed formulas such as Eqs. $\left( \ref{fsl}\right) $ and $\left( 
\ref{fal}\right) $, which can be readily used in order to correct the cross
section measurements. Moreover, the latter model compares well with other
available theories and its use is expected to be particularly useful in any
future attempt to improve the accuracy of the $CNO$ astrophysical factors by
lowering the energy of the experiment.

Finally, the fitting method used in the determination of the astrophysical
factor is identified here as a source of uncertainty for the solar neutrino
fluxes, which needs further elaboration and justification.

\section{Appendix}

The screened potential model approach needs some elaboration regarding its
actual effects. In Refs. \cite{rolfsbook} and \cite{assen} the screening
energy for a collision between the atomic target $\left( Z_{1}e\right) $ and
the projectile $\left( Z_{2}e\right) $ was identified as

\begin{equation}
U_{e}=\frac{Z_{1}Z_{2}e^{2}}{R_{a}}  \label{uera}
\end{equation}
where the screening radius was set equal to the radius of the innermost
electrons of the target $\left( R_{a}=a_{0}/Z_{1}\right) \,$labeling that as
the ''worst case''. In that model, the electron cloud is assumed to be
unperturbed , which is the definition for the sudden limit, and there is no
special consideration for multi-electron effects, either.

However, in our study of hydrogen-like atomic targets\cite{lioliosepj} we
have shown that the screening radius $R_{a}$, for the same sudden limit, is
actually independent of the atomic number $Z_{1}$ and equal to the Bohr
radius $R_{a}=a_{0}.$, thus obtaining a smaller screening energy when $%
Z_{1}>1$ than the one used in Refs. \cite{rolfsbook} and \cite{assen}. Note
that if we adopt the ''worst case'' approach for the reaction $^{12}C\left(
p,\gamma \right) ^{13}N,$ then Eq. $\left( \ref{uera}\right) $ gives the
unrealistic screening energy $U_{e}=1000$ $eV$ which is beyond the present
TF adiabatic limit value, too.

On the contrary, it is easy to show our approximation is valid for most
practical purposes. The screening term $U_{e}^{\left( r\right) }$ that we
disregarded when treating Eq. $\left( \ref{v12}\right) $ via the WKB is
actually proportional to the scaled relative internuclear distance $r/a_{0}:$

\begin{equation}
U_{e}^{\left( r\right) }=-2Z_{1}^{2}Z_{2}\frac{e^{2}}{a_{0}}\frac{r}{a_{0}}
\label{ue2}
\end{equation}
For a typical classical turning point $r_{c}\sim 10^{-2}a_{0}$ we have,
throughout the barrier, $r<10^{-2}a_{0}$. Hence, in the sudden limit, the
ratio of the term that we considered significant $U_{e}$ (Eq. $\left( \ref
{usl}\right) $) to the insignificant one, given by Eq. $\left( \ref{ue2}%
\right) ,$ is $U_{e}/U_{e}^{\left( r\right) }>50Z_{1}^{-1}.$ Obviously for
small $Z_{1}$ , which is usually the case in astrophysical reactions, our
formulas becomes increasingly accurate.

{\bf ACKNOWLEDGMENTS}

This work was financially supported by the Greek State Grants Foundation
(IKY) under contract \#135/2000. It was initiated at ECT$^{*}$ during a
nuclear physics fellowship. The author would like to thank the director of
ECT$^{*}$ Prof.Malfliet for his kind hospitality and support.

\bigskip FIGURE CAPTIONS

Figure 1. The screening enhancement factor $f_{TF}\left( E\right) $ for the
most important astrophysical nuclear reactions of the $CNO$ bi-cycle with
respect to the center-of-mass energy $E$ in the region of the Gamow peaks $%
\left( E_{GP}\right) $ as calculated for central solar conditions. The lower
(upper) solid curve represents the enhancement of the $^{13}C\left( p,\gamma
\right) ^{14}N$ reaction $\left( E_{GP}=24.5\text{ }keV\right) $ as
calculated by the above TF sudden (adiabatic) limit. Likewise, the dashed
curves stand for the $^{14}N\left( p,\gamma \right) ^{15}O$ reaction $\left(
E_{GP}=27.2\text{ }keV\right) $, while the dotted ones for the $^{16}O\left(
p,\gamma \right) ^{17}F$ reaction $\left( E_{GP}=29.8\text{ }keV\right) .$
Lower and upper curves always indicate sudden and adiabatic limits
respectively.

Figure 2. The screening enhancement factor $f_{TF}\left( E\right) $ for the
most important astrophysical nuclear reactions of the $CNO$ bi-cycle with
respect to the center-of-mass energy in the region $2\leq E\leq 5keV$. The
enhancement effect is particularly accentuated at such low energies. The
notation is the same as in Fig.1.


\begin{references}
\bibitem{lunasecond}  R.Bonetti et al, Phys.Rev.Lett. {\bf 82} 5205(1999)

\bibitem{bahcallbook}  J.N.Bahcall, 1989, Neutrino Astrophysics, Cambridge
University Press

\bibitem{kraussdd}  A.Krauss, H.W. Becker, H.P.Trautvetter, C.Rolfs,
Nucl.Phys.A{\bf 465}, 150 $(1987)$

\bibitem{krausshe3}  A.Krauss, H.W. Becker, H.P.Trautvetter, C.Rolfs,
Nucl.Phys.A{\bf 467} 273 $(1987)$

\bibitem{greife}  U.Greife, F.Gorris, M.Junker, C.Rolfs, D.Zahnow, Z.Phys.A.
351, 107$\left( 1995\right) $

\bibitem{brown}  R.E.Brown, N.Jarmie, Phys.Rev.C{\bf 41}, 1391$\left(
1990\right) $

\bibitem{engstler}  S.Engstler, A.Krauss, K.Neldner, C.Rolfs, U.Schroeder,
Phys.Lett.B. 202, 179$\left( 1998\right) $

\bibitem{liolios2000}  T.E.Liolios, Proceedings of the Bologna 2000
Conference: Structure of the Nucleus at the Dawn of the Century, Bologna $%
\left( 2000\right) $ (in press)

\bibitem{rolfs2000}  C.Rolfs, Proceedings of the Bologna 2000 Conference:
Structure of the Nucleus at the Dawn of the Century, Bologna $\left(
2000\right) $ (in press)

\bibitem{rolfsbook}  C.E.Rolfs, W.S.Rodney, ''Cauldrons in the Cosmos'' ,
The University of Chicago Press, (1988)

\bibitem{shoppamolecular}  T.D.Shoppa, M.Jenk, S.E.Koonin, K.Langanke,
R.Seki, Nucl. Phys.A605 387$(1996$)

\bibitem{bencze}  G.Bencze, Nucl.Phys.A{\bf 492}, 459 $\left( 1989\right) $

\bibitem{bang}  J.M.Bank, L.S.Ferreira, E.Maglione, J.M.Hansteen, Phys.Rev.%
{\bf C.53}, R18$\left( 1996\right) $

\bibitem{langloss}  K.Langanke, T.D.Shoppa, C.A .Barnes, C.E.Rolfs, Phys.
Lett. {\bf B369} 211$(1996)$

\bibitem{assen}  H.J.Assebaum, K.Langanke, C.Rolfs, Z.Phys.A {\bf 327}, 461 $%
\left( 1987\right) $

\bibitem{bracci}  L.Bracci, G.Fiorentini, V.S.Melezhik, G.Mezzorani,
P.Quarati, Nucl.Phys.A{\bf 513}, 316$\left( 1990\right) $

\bibitem{lioliosepj}  T.E.Liolios, nucl-th/0005011, Eur.Phys.J.A $\left(
2000\right) $ in press

\bibitem{lunafirst}  M.Junker et al, Phys.Rev. {\bf C57} 2700(1998)

\bibitem{gazbook}  S. Gasiorowicz, ''Quantum Physics'', John Wiley \& Sons,
Inc., $\left( 1996\right) $

\bibitem{flugge}  S.Flugge, ''Practical Quantum Mechanics'', Springer-Verlag 
$\left( 1998\right) $ ISBN \#3540650350

\bibitem{tietz}  T.Tietz, J.Chem.Physics ${\bf 25}$, 787$\left( 1956\right) $

\bibitem{shoppaatomic}  T.D.Shoppa, S.E.Koonin, K.Langanke, R.Seki,
Phys.Rev.C{\bf 48}, 837$\left( 1990\right) $

\bibitem{lioliosprc}  T.E.Liolios, Phys.Rev.C 61, 55802$\left( 2000\right) $

\bibitem{azuma}  C.Rolfs, R.E.Azuma, Nucl.Phys.{\bf A227}, 292$\left(
1974\right) $

\bibitem{schroeder}  U.Schroeder, H.W.Becker, G.Bogaert, J.Goerres, C.Rolfs,
H.P.Trautvetter, Nucl.Phys.{\bf A476}, 240$\left( 1987\right) $
\end{references}
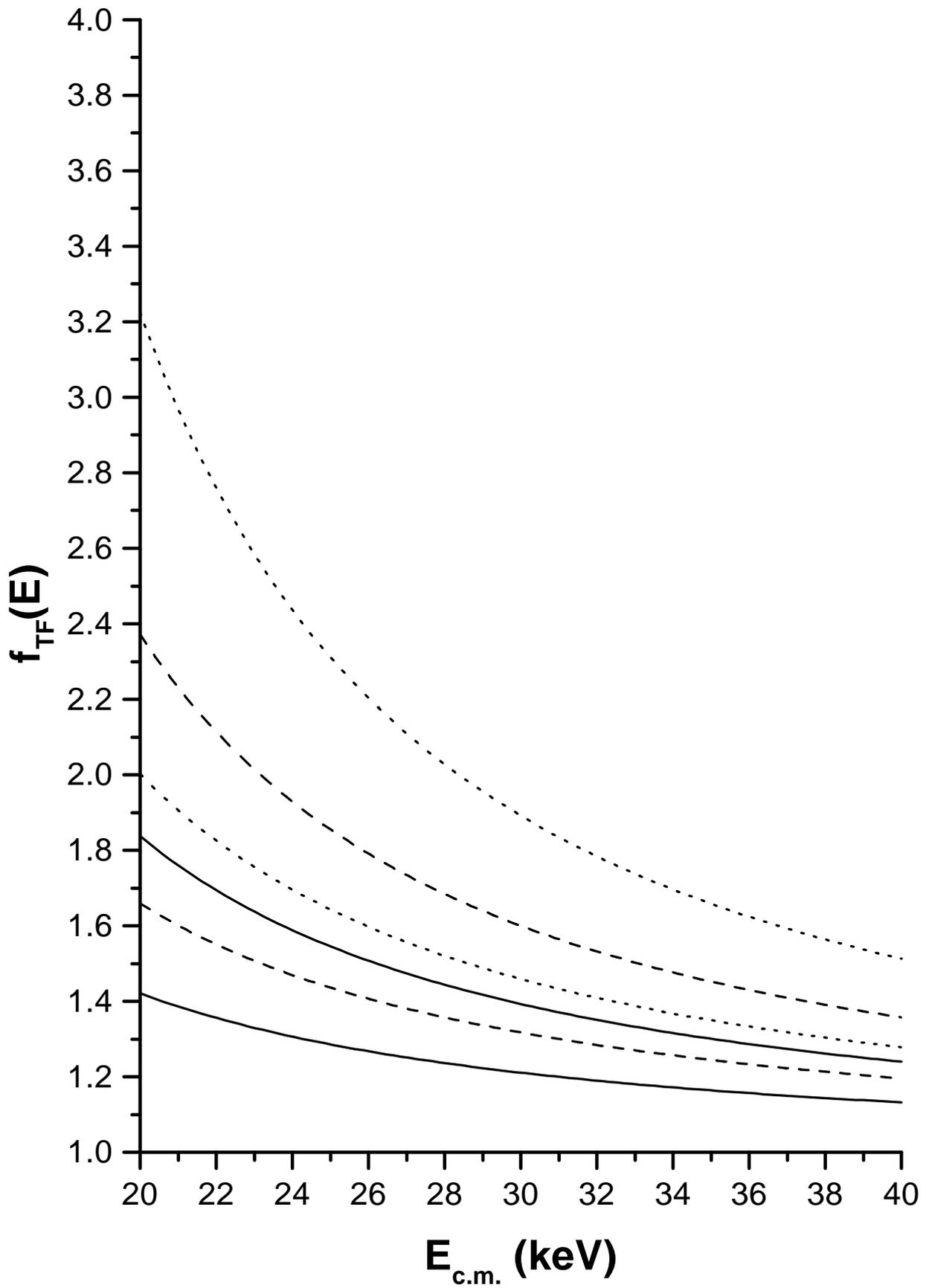
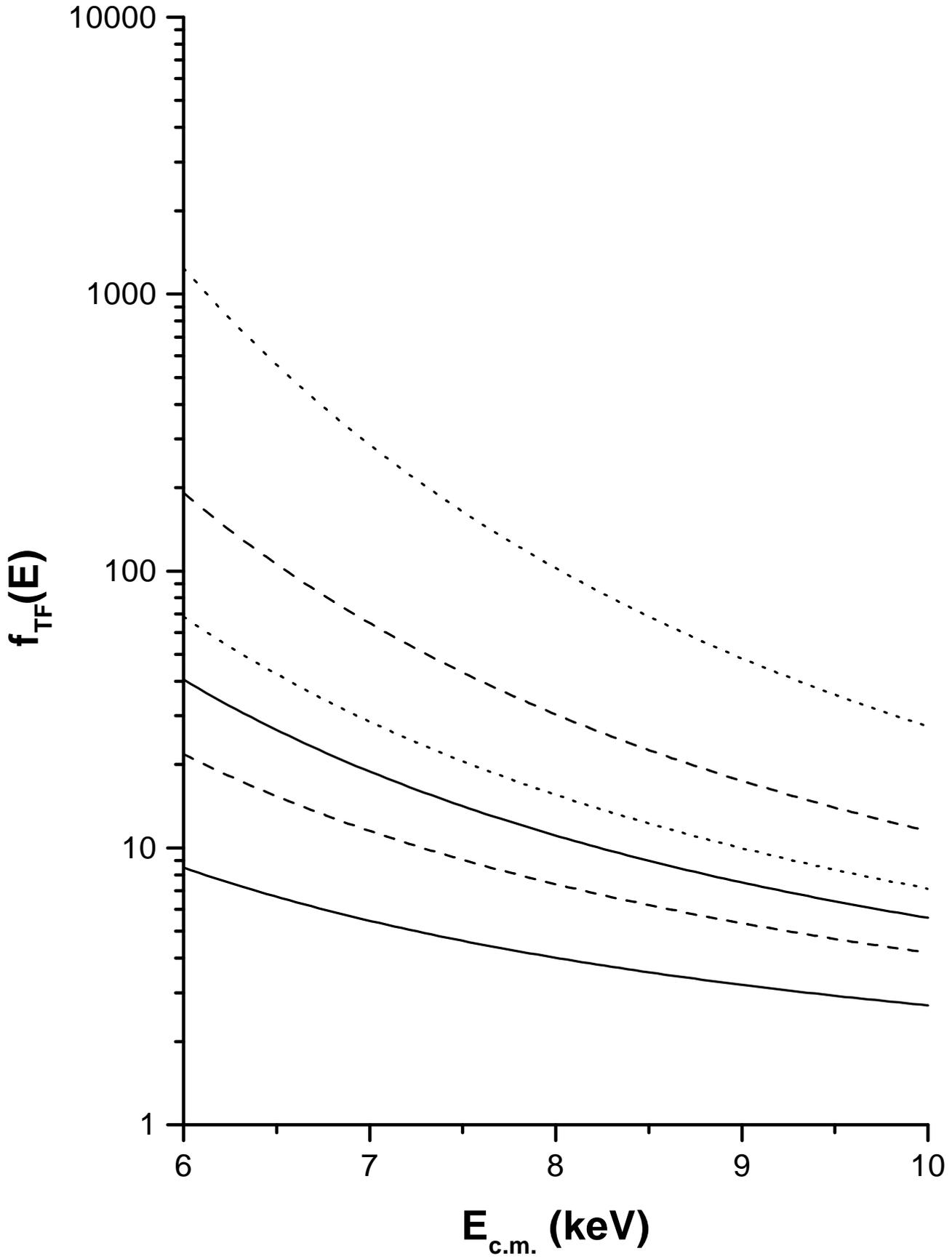
\end{document}